# High energetic excitons in carbon nanotubes directly probe charge-carriers

Giancarlo Soavi[1], Francesco Scotognella,[1,2,3], Daniele Viola[1], Timo Hefner[4], Tobias Hertel[4], Giulio Cerullo [1, 2]* and Guglielmo Lanzani[1, 3]†

[1] *Dipartimento di Fisica, Politecnico di Milano, Piazza L. da Vinci 32, 20133 Milano, Italy*

[2] *IFN-CNR, Piazza L. da Vinci, 32, 20133 Milano, Italy*

[3] *Center for Nano Science and Technology@PoliMi, Istituto Italiano di Tecnologia, Via Giovanni Pascoli, 70/3, 20133 Milano, Italy*

[4] *Inst. for Physical and Theoretical Chemistry Dept. of Chemistry and Pharmacy, University of Wuerzburg, Wuerzburg 97074, Germany*

\* giulio.cerullo@polimi.it, † guglielmo.lanzani@iit.it

**Abstract**

Theory predicts peculiar features for excited-state dynamics in one dimension (1D) that are difficult to be observed experimentally. Single-walled carbon nanotubes (SWNTs) are an excellent approximation to 1D quantum confinement, due to their very high aspect ratio and low density of defects. Here we use ultrafast optical spectroscopy to probe photogenerated charge-carriers in (6,5) semiconducting SWNTs. We identify the transient energy shift of the highly polarizable $S_{33}$ transition as a sensitive fingerprint of charge-carriers in SWNTs. By measuring the coherent phonon amplitude profile we obtain a precise estimate of the Stark-shift and discuss the binding energy of the $S_{33}$ excitonic transition. From this, we infer that charge-carriers are formed instantaneously with sizable quantum yield even upon pumping the first exciton, $S_{11}$. The decay of the photogenerated charge-carrier population is well described by a model for geminate recombination in 1D, suggesting an initial charge carrier separation of the same order of the exciton correlation length.

The study of photo-excitation dynamics in one dimension has been prompted by theoretical predictions of a wealth of singular properties, such as the giant oscillator strength and non-linear response of confined states, the large Coulomb interaction, the sharply-peaked density of states and peculiar excited-state recombination kinetics [1-3]. In this respect, SWNTs represent a very close approximation to a 1D solid, easily achieving aspect ratio as high as $10^3$. Theory predicts that Wannier-Mott excitons are the elementary photoexcitations in SWNTs, due to the strong Coulomb interaction caused by limited screening [4, 5]. These excitons have typical 1D characteristics: negligible free carrier generation, large binding energy, non-negligible size and 1D transport. Theoretical predictions are supported by several experimental results, such as the measured binding energy, typically 0.1-1 eV [6, 7], and the electron-hole correlation length, in the 2-5 nm range [8]. The exciton model alone, however, fails to capture the whole dynamics following photoexcitation, and many other photoexcited species have crowded the complex scenario of SWNTs' optical response, ranging from triplets [9] to bi-excitons [10] and trions [11]. Photocurrent [12-16], transient absorption [14, 17, 18] and THz spectroscopy [19, 20] experiments also point out a non-negligible photogeneration of free charge-carriers in SWNTs. This is in stark contrast with the excitonic model and the reduced Sommerfeld factor that implies excitons be the only species generated upon photoexcitation. Attempts to solve this discrepancy proposed possible non-linear phenomena [21] as the mechanism of charge-carrier photogeneration in SWNTs. However, there is solid experimental evidence that the charge-carrier yield is linear with the pump fluence [19]. Besides this, the nature of high energetic transitions in small-diameter semiconducting SWNTs is still matter of debate, given that both excitonic [22, 23] and band-to-band transitions [24, 25] have been invoked to explain recent experimental results.

In this Letter we apply ultrafast optical spectroscopy to the semiconducting (6,5) SWNTs and show that charge-carriers can be identified by their effect on excitonic resonances, in particular the large energy shift that they induce on the third excitonic subband ($S_{33}$) transition. The availability of a good fingerprint for charge-carriers enables us to study their dynamics in one dimension. We find that, upon excitation of the lowest optical transition, a fraction of the absorbed photons generates geminate charge-carrier pairs "instantaneously". The carriers recombine on the sub-nanosecond timescale following the characteristic kinetic law ($\sim t^{-1/2}$) of a random walk in 1D, which suggests that the initial charge carrier separation is of the same order as the exciton correlation length.

The sample used for these investigations is highly enriched in the (6,5) species and embedded in a gelatin film. This film was prepared from 30 microliters of a density gradient ultracentrifugation (DGU) enriched SWNT suspension in a sodium cholate (SC)/sodium dodecyl sulfate (SDS) mixture [26]. Iodixanol as well as SDS residues from the DGU process were removed by dilution with SC solution and filtration with a benchtop centrifuge. The resulting suspension with 30 microliters volume was then mixed with 20 microliters of 15 wt% gelatin solution and finally drop-cast on a thin glass substrate. Ultrafast pump-probe spectroscopy was carried out on a very broad wavelength region from 340 nm to 1.1 µm, thus probing the transient absorption signal of the third ($S_{33}$), second ($S_{22}$) and first ($S_{11}$) excitonic transitions of the sample (Fig. 1b). We excited the sample either with a broad IR pulse, peaked around 1 µm and with a transform-limited pulse duration of less than 15 fs (for measurements in Fig. 2 and 3) or with a 10-nm bandwidth pulse peaked at 570 nm (for measurements in Fig. 1b). As a probe we used: i) the second harmonic of a visible optical parametric amplifier (OPA), in order to achieve an overall temporal resolution of ≈50 fs in the probe region from 340 nm to 370 nm; ii) broadband white light super-continuum generated in $CaF_2$ in the probe region from 340 nm to 650 nm and iii) broadband white light super-continuum generated in a sapphire plate in the probe region from 850 nm to 1.1 µm. We measured the differential transmission ($\Delta T/T$) through the sample with an optical multichannel analyzer working at the full repetition rate (1kHz) of the laser source [27].

Figure 1a shows the linear absorption spectrum of the sample with its first three excitonic transitions: $S_{11}$ near 1 µm, $S_{22}$ near 570 nm and $S_{33}$ near 350 nm. Figure 1b shows $\Delta T/T$ spectra for 570 nm excitation wavelength at different pump-probe delays. In agreement with our previous work [14], we find that the shape of the transient spectral response does not depend on the excitation wavelength (Supplementary material). We observe three sharp positive $\Delta T/T$ peaks corresponding to the three excitonic transitions, each associated with negative features, smaller than the positive one for the first and second excitons, but comparable for the third exciton. The large positive peak in the first exciton region can be assigned to photobleaching (PB) due to state filling. The photoinduced absorption (PA) above 1.1 µm has been tentatively assigned to triplets [28], metallic tubes [29], trions [21], transitions from $S_{11}$ to the first band edge [30] or bi-excitons [31]. The complex shape of the $\Delta T/T$ signal around $S_{22}$ can be reproduced with a red shift or a broadening of the ground state absorption spectrum. Several processes, such as photoinduced dephasing [32], bi-exciton formation [33, 34], phonon dynamics [35] and charge induced Stark effect [14, 18] have been invoked to explain the transient signal in this spectral region. Both the first and second excitons thus present complex transients, due to the superposition of several overlapping contributions. On the other hand, the third excitonic sub-band shows a simple first derivative lineshape that corresponds to a photoinduced red shift of the ground state transition.

Figure 2 zooms in on the transient $\Delta T/T$ spectra and dynamics in the region near $S_{33}$ when the sample is excited at the $S_{11}$ transition, with ≈50-fs temporal resolution (Fig. 2b). After rapid initial changes in the first ≈150 fs (Fig. 2a), the shape of the transient spectra remains unvaried up to 1 ns, the longest delay investigated here (inset of Fig. 2a). The observed derivative shape is insensitive to the pump-photon energy (Supplementary material), thus excluding bi-excitons and trions. Intensity dependent measurements (Fig. 2c and inset) demonstrate that the experiments are performed in the linear regime with respect to the pump pulse, thus ruling out non-linear processes such as two-photon absorption or exciton-exciton annihilation [21], which is expected to occur in a saturation regime for exciton photogeneration [36, 37]. Similarly, the signal is weakly sensitive to changes in temperature (Supplementary material), excluding geometrical re-arrangement (i.e. diameter distortion) and thermal effects as a possible origin of the strong red-shift of $S_{33}$ upon photoexcitation. In order to better understand the origin of the $\Delta T/T$ signal for $S_{33}$, we fit it by the sum of three contributions (Fig. 2d and Supplementary material): a Lorentzian function, corresponding to the ground-state PB, the difference between two Lorentzian functions, corresponding to a spectral red-shift by 0.13 eV, and a constant PA over the entire probe bandwidth. We conclude that the ΔT/T signal for the third exciton is dominated by the spectral shift, that we assign to the Stark shift induced on the $S_{33}$ transition by the intense local electric field of photogenerated charge-carriers. The Stark effect depends on the transition's polarizability and is enhanced for excitons with small binding energy, such as $S_{33}$ [24]. Note that, in contrast to state filling which selectively affects only transitions involving populated states, the Stark effect lacks this selectivity and affects all optical transitions. Our assignment is based on the following chain of reasoning: i) charge-carriers are photogenerated in SWNTs; ii) each charge-carrier is a source of a strong local electric fields; iii) the amplitude of the Stark signal has a distinct kinetics from that of the exciton PB, being in particular much longer lived; iv) any other source of modulation that could explain the first derivative shape of the ΔT/T spectra for $S_{33}$ has been ruled out. Furthermore, the observation of a first derivative shape only for the highly polarizable $S_{33}$ transition is readily accounted for by the Stark effect, while it could not be justified for other modulations that should affect all excitons in the same way. We propose that photo-excitation at the $S_{11}$ transition creates both excitons and free charge-carriers: the first bleach $S_{33}$, the latter shift the energy levels due to Stark effect. The Stark effect prevails, making the $S_{33}$ transition privileged to probe charge-carriers, for at least two reasons: i) the $S_{33}$ exciton has small cross-section and thus small PB signal; ii) it has low binding energy [24, 25], resulting in large field-induced energy shift. The fit indicates that PB, PA and Stark shift are all formed within our temporal resolution, as confirmed by the ultrafast build-up of the ΔT/T signal (Fig. 2b). This is in good agreement with the claim of instantaneous charge photogeneration of ref. 20 and only slightly differs from the results obtained in ref. 19, where the rate of linear exciton dissociation is ≈ 3.4 ps$^{-1}$. This discrepancy is likely due to the limited temporal resolution available in THz experiments (≈ 1 ps). The PB signal decays faster (≈ 600 fs) with respect to the Stark signal, as expected for the lifetime of excitons with respect to free or trapped charge-carriers. The fast decay of the PB signal also suggests that at longer delays (i.e. few picoseconds to nanoseconds) the transient signal at the $S_{33}$ transition directly probes charge-carriers.

So far, we have assumed that the $S_{33}$ transition is excitonic in nature and it has low binding energy. Nevertheless, the nature of $S_{33}$ in semiconducting SWNTs is still matter of debate. Rayleigh scattering experiments have demonstrated that it is consistent with an excitonic model [22], theoretical studies predict high binding energies [23] while recent experiments shows features ascribable to unbound electron-hole pairs [24, 25]. The PB contribution to the $S_{33}$ transient signal (Fig. 2d) suggests that this transition is excitonic in nature. To study its binding energy in comparison with the lower energy excitonic transitions, we study the clear periodic temporal modulations of the ΔT/T signal around the $S_{33}$ transition (Fig. 3a), that we assign to impulsively excited coherent phonons, namely the radial breathing modes (RBMs). Their coupling to the optical response of SWNTs can be easily understood: exciton binding energies are approximately inversely proportional to the tube diameter [38, 39], so that the exciton absorption peak undergoes red or blue shift according to diameter variations. The resulting oscillations have zero amplitude at the peak of the resonance and maximum amplitude with opposite phase for higher and lower energies [40]. Figure 3b shows the oscillatory component of the $\Delta T/T$ signal for two wavelengths near the $S_{33}$ peak. Fourier transform of the time traces indicates a dominant frequency of 318 cm$^{-1}$, consistent with the RBM of the (6,5) SWNT. Figure 3a shows the corresponding amplitude and phase of the modulation versus probe energy. Interestingly, the peak resonance energy, indicated by the zero amplitude and by the phase jump (Fig. 3a), is red-shifted from the ground state S$_{33}$ transition (at 349 nm). In other words, coherent RBM phonons modulate the Stark-shifted transition, thus providing a very sensitive tool for measuring the Stark shift of the exciton resonance. We obtain $\Delta E_{Stark} \sim 130\ meV$, in excellent agreement with the results of the fitting model. The quadratic part of the Stark effect in SWNTs can be expressed as $\delta E_{Stark} = \kappa_b (edF^2)/E_b$, where $\kappa_b$ is a fitting constant, $e$ is the electron charge, $d$ the SWNT diameter, $F$ the electric field responsible for the Stark shift and $E_b$ the exciton binding energy [41]. Since the Stark effect is inversely proportional to the exciton binding energy, we conclude that the large red shift that we observe on the S$_{33}$ exciton is a direct consequence of its low binding energy (in agreement with Ref. 24), in particular if compared to lower-energy excitons (S$_{11}$ and S$_{22}$), where the Stark shift is considerably smaller.

Our data also give us the unique opportunity to analyze the dynamics of free charge-carriers in one dimension. Due to the linear dependence of the $\Delta T/T$ signal on pump fluence (inset Fig. 2c), the most probable mechanisms of charge photogeneration remain either direct excitation or ultrafast linear exciton dissociation. According to our analysis, the charge-carrier population decay is monitored by the $\Delta T/T$ time trace at 363 nm, which represents the Stark amplitude evolution. Figure 3a shows that this decay is very accurately reproduced by a power law $(\Delta T/T)_{norm} = A \cdot t^{-1/2}$. The same power law is observed at different pump intensities (Supplementary material), thus excluding that the non-exponential kinetics is due to bimolecular non-geminate carrier annihilation. A monomolecular power law decay is the predicted dynamics for geminate recombination of free particles after random walk in an infinite one-dimensional chain [42]. In particular, the probability $\Omega(t)$ to survive geminate recombination at delay $t$ is expressed by $\Omega(t) = n_0/\sqrt{2\pi W t}$, where $W$ is the diffusion rate and $n_0 = L_0/a_0$ is the normalized initial particle separation, $L_0$ being the particle distance and $a_0$ the unit cell. Considering that a hopping model is a good approximation for the high-mobility diffusive transport of SWNTs [44, 45], we estimate the diffusion rate W as the inverse of the scattering time $\tau = m_{av}\mu/(0.32e)$, where $m_{av}$ is the effective mass and $\mu$ is the charge-carrier mobility [43]. Using the value for mobility in a (6,5) SWNT of $\mu \sim 10^3$ (diameter of $\approx$ 1nm) and assuming $a_0 \sim 1\,nm$, we obtain $L_0 \sim 5\,nm$, meaning that the initial distance between a geminate electron-hole pair is of the same order of magnitude of the excitonic correlation distance [8]. This suggests that "instantaneous" (within 50-fs) linear exciton dissociation is the most likely mechanism of charge photogeneration. Possibly this process is favoured by the presence of atmospheric contamination, due to water and/or oxygen, which strongly reduces the exciton binding energy [45-49]. A similar power-law kinetics was already observed in ensembles of semiconducting SWNTs for high fluence excitation on the $S_{11}$ excitonic transition [44, 50]. This was assigned to either bimolecular triplet or singlet annihilation, in contrast with our work that shows, for the first time, a charge diffusion limited geminate recombination in a single chirality specimen and in the linear excitation regime.

In conclusion, we identified the transient energy shift of the $S_{33}$ transition as a sensitive fingerprint of charge-carriers in (6,5) SWNTs. The assignment is based on the notion that photogenerated charge-carriers give rise to strong electric fields that in turn shift the energy of the more polarisable transition, namely the $S_{33}$ resonance. Modulation spectroscopy through coherent phonons allows a precise measurement of the Stark shift of the $S_{33}$ exciton, from which we estimate its binding energy in comparison with lower energy excitonic transitions. Our results indicate that charge-carriers are formed with sizable quantum yield even upon pumping the first exciton, $S_{11}$. This is a surprising outcome, since in principle $S_{11}$ states are strongly bound and located well below the continuum. The decay of the photogenerated charge-carrier population is well described by a model for geminate recombination in 1D. From the model we estimate an initial charge carrier separation of the same order of the exciton correlation length. This sheds additional light onto the generation mechanism, suggesting that the nascent excitons dissociate spontaneously, perhaps in presence of extrinsic screening of the Coulomb attraction due to water or other contamination. This result implies that charge photogeneration in SWNTs can be engineered, for instance for applications in optoelectronics, by manipulation of the tube environment. Long-lived charge carriers can only be obtained by promoting intertube separation, to escape efficient geminate recombination.

**Acknowledgment**

C.G. acknowledges support by the EC under Graphene Flagship (contract no. CNECT-ICT-604391). F.S., G.L. and T.H. acknowledge the ITN project 316633 "POCAONTAS".

**References**

[1] G. D. Scholes and G. Rumbles, Nat. Mater. **5**, 683 (2006).

[2] G. D. Scholes, ACS nano **2**, 523 (2008).

[3] J. H. Davies *The physics of low-dimensional semiconductors* (Cambridge University Press: Cambridge, 1998).

[4] T. Ando, J. Phys. Soc. Jpn. **66**, 1066 (1997).

[5] M. Ichida, S. Mizuno, Y. Tani, Y. Saito and A. Nakamura, J. Phys. Soc. Japan **68,** 3131 (1999).

[6] F. Wang, G. Dukovic, L. E. Brus and T. F. Heinz, Science **308**, 838 (2005).

[7] J. Maultzsch, R. Pomraenke, S. Reich, E. Chang, D. Prezzi, A. Ruini, E. Molinari, M. S. Strano, C. Thomsen and C. Lienau, Phys. Rev. B **72**, 241402 (2005).

[8] L. Lüer, S. Hoseinkhani, D. Polli, J. Crochet, T. Hertel and G. Lanzani, Nat. Phys. **5**, 54 (2008).

[9] D. Stich, F. Späth, H. Kraus, A. Sperlich, V. Dyakonov and T. Hertel, Nat. Photonics **8**, 139 (2014).

[10] T. G. Pedersen, K. Pedersen, H. D. Cornean and P. Duclos, Nano Lett. **5**, 291 (2005).

[11] T. F. Rønnow, T. G. Pedersen and H. D. Cornean, Phys. Rev. B **81**, 205446 (2010).

[12] A. Malapanis, V. Perebeinos, D. P. Sinha, E. Comfort and J. U. Lee, Nano Lett. **13**, 3531 (2013).

[13] D. J. Bindl, A. J. Ferguson, M. Wu, N. Kopidakis, J. L. Blackburn, M. S. Arnold, M. S. J. Phys. Chem. Lett. **4,** 3550 (2013).

[14] G. Soavi, F. Scotognella, D. Brida, T. Hefner, F. Späth, M. R. Antognazza, T. Hertel, G. Lanzani and G. Cerullo, J. Phys. Chem. C **117**, 10849 (2013).

[15] Y. Kumamoto, M. Yoshida, A. Ishii, A. Yokoyama, T. Shimada and Y. K. Kato, Phys. Rev. Lett. **112**, 117401 (2013).

[16] M. Barkelid, G. A. Steele and V. Zwiller, Nano Lett. **12**, 5649 (2012).

[17] C. Gadermaier, E. Menna, M. Meneghetti, W. J. Kennedy, Z. V. Vardeny and G. Lanzani, Nano Lett. **6**, 301 (2006).

[18] J. Crochet, S. Hoseinkhani, L. Lüer, T. Hertel, S. K. Doorn and G. Lanzani, G. Phys. Rev. Lett. **107**, 257402 (2011).

[19] M. C. Beard, J. L. Blackburn and M. J. Heben, Nano Lett. **8**, 4238 (2008).

[20] S. A. Jensen, R. Ulbricht, A. Narita, X. Feng, K. Müllen, T. Hertel, D. Turchinovich and M. Bonn, Nano Lett. **13**, 5925 (2013).

[21] S. M. Santos, B. Yuma, S. Berciaud, J. Shaver, M. Gallart, P. Gilliot, L. Cognet, and B. Lounis, Phys. Rev. Lett. **107**, 187401 (2011).

[22] S. Berciaud, C. Voisin, H. Yan, B. Chandra, R. Caldwell, Y. Shan, L. E. Brus, J. Hone and T. F. Heinz, Phys. Rev. B **81**, 041414 (2010).

[23] C. D. Spataru, S. Ismail-Beigi, L. X. Benedict and S. G. Louie, Phys. Rev. Lett. **92**, 077402 (2004).

[24] P. T. Araujo, S. K. Doorn, S. Kilina, S. Tretiak, E. Einarsson, S. Maruyama, H. Chacham, M. A. Pimenta and A. Jorio, Phys. Rev. Lett. **98**, 067401 (2007).

[25] T. Michel, M. Paillet, J. C. Meyer, N. V. Popov, L. Henrard and J. L. Sauvajol, Phys. Rev. B **75**, 155432 (2007).

[26] J. Crochet, M. Clemens and T. Hertel, J. Am. Chem. Soc. **129**, 8058 (2007)

[27] D. Polli, L. Lüer and G. Cerullo, Rev. Sci. Intrum. **78**, 103108 (2007)

[28] J. Park, P. Deria and M. J. Therien, J. Am. Chem. Soc. **133**, 17156 (2011).

[29] L. Lüer, G. Lanzani, J. Crochet, T. Hertel, J. Holt and Z. V. Vardeny, Phys. Rev. B **80**, 205411 (2008).

[30] Y.-Z. Ma, L. Valkunas, S. M. Bachilo and G. R. Fleming, J. Phys. Chem. B **109**, 15671 (2005).

[31] J. Park, P. Deria, J.-H. Olivier and M. J. Therien, Nano Lett. **14**, 504 (2014).

[32] G. Ostojic, S. Zaric, J. Kono, V. C. Moore, R. H. Hauge and R. E. Smalley Phys. Rev. Lett. **94**, 097401 (2005).

[33] D. J. Styers-Barnett, S. P. Ellison, B. P. Mehl, B. C. Westlake, R. L. House, C. Park, K. E. Wise and J. M. Papanikolas, J. Phys. Chem. C **112**, 4507 (2008).

[34] B. Gao, G. V. Hartland and L. Huang, ACS nano **6**, 5083 (2012).

[35] T. Koyama, S. Yoshimitsu, Y. Miyata, H. Shinohara, H. Kishida and A. Nakamura, J. Phys. Chem. C **117**, 20289 (2013).

[36] Y.-F. Xiao, T. Q. Nhan, M. W. B. Wilson and J. M. Fraser, Phys. Rev. Lett. **104**, 017401 (2010).

[37] J. R. Schneck, A. G. Walsh, A. A. Green, M. C. Hersam, L. D. Ziegler and A. K. Swan, J. Phys. Chem. A **115**, 3917 (2011).

[38] H. Kataura, Y. Kumazawa, Y. Maniwa, I. Umezu, S. Suzuki, Y. Ohtsuka and Y. Achiba, Synth. Met. **103**, 2555 (1999).

[39] V. Perebeinos, J. Tersoff and P. Avouris, Phys. Rev. Lett. **92,** 257402 (2004).

[40] Y.-S. Lim, K. Yee, J. Kim, E. H. Hároz, J. Shaver, J. Kono, S. K. Doorn, R. H. Hauge and R. E. Smalley, Nano Lett. **6**, 2696 (2006).

[41] V. Perebeinos and P. Avouris, Nano Lett. **7**, 609 (2007).

[42] I. V. Zozulenko, Solid State Commun. **76**, 1035 (1990).

[43] X. Zhou, J.-Y. Park, S. Huang, J. Liu and P. McEuen, Phys. Rev. Lett. **95**, 146805 (2005)

[44] R. Russo, E. Mele, C. Kane, I. Rubtsov, M. Therien and D. Luzzi, Phys. Rev. B **74**, 041405(R) (2006).

[45] J. S. Bulmer, J. Martens, L. Kurzepa, T. Gizewski, M. Egilmez, M. G. Blamire, N. Yahya and K. K. K. Koziol, Sci. Rep. **4**, 3762 (2014).

[46] J. Lefebvre and P. Finnie, Nano Lett. **8**, 1890 (2008).

[47] Q. Zhang, E. H. Hároz, Z. Jin, L. Ren, X. Wang, R. S. Arvidson, A. Lüttge and J. Kono Nano Lett. **13**, 5991 (2013).

[48] A. Zahab, L. Spina, P. Poncharal and C. Marlière, Phys. Rev. B **62**, 10000 (2000).

[49] D. Kang, N. Park, J. Ko, E. Bae and W. Park, Nanotechnology **16**, 1048 (2005).

[50] P. G. Collins, K. Bradley, M. Ishigami and A. Zettl, Science **287**, 1801 (2000).

[51] J. Allam, M. T. Sajjad, R. Sutton, K. Litvinenko, Z. Wang, S. Siddique, H. Q. Yang, W. H. Loh and T. Brown, Phys. Rev. Lett **111**, 197401 (2013).

**Figure Captions**

**Figure 1.** (color online) Linear and transient absorption spectra of SWNTs. a) Absorption spectrum of the enriched (6,5) SWNTs sample. b) Transient absorption spectra at different pump-probe delays, with 570 nm excitation wavelength, for a (6,5) enriched SWNT sample. The probe is obtained with white light continuum, from $CaF_2$ for the wavelengths from 340 nm to 650 nm and from sapphire for wavelengths from 850 nm to 1100 nm.

**Figure 2.** (color online) (a) $\Delta T/T$ spectra at different pump-probe delays (solid line) and $\Delta T/T$ spectra obtained from the fitting (triangles). The excitation fluence is approximately 100 µJ/cm$^2$ for the main figure and 200 µJ/cm$^2$ for the inset. (b) Dynamics for wavelengths on the positive and negative peaks of the signal compared to our temporal resolution (≈ 50 fs). The pump pulse excites the first excitonic subband $S_{11}$ in the IR region (≈ 1µm). (c) Transient spectra at different pump intensities for 150-fs pump-probe delay and (inset) absolute value of the $\Delta T/T$ signal at 363 nm as a function of the pump fluence. (d) Spectral components used for the fitting, i.e. PB, PA and Stark shift, at 500fs pump-probe delay.

**Figure 3.** (color online) (a) Fourier transform's amplitude and phase for the (6,5) RBMs' frequency (318 cm$^{-1}$) near the $S_{33}$ energy region. (b) Oscillatory component at 357 nm (blue) and 372 nm (red), the two maxima of RBMs oscillations' amplitude. (c) Normalized dynamics at 363 nm (black) and fit (red) with a $A/\sqrt{t}$ function, where $t$ is time and $A$ is a fitting parameter. The excitation fluence is approximately 90 µJ/cm$^2$.

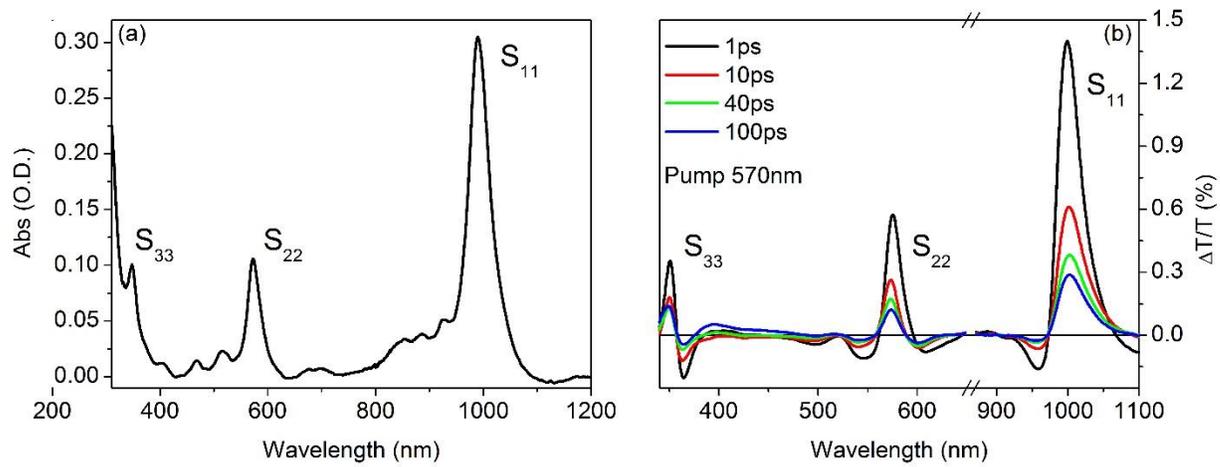

Figure 1.

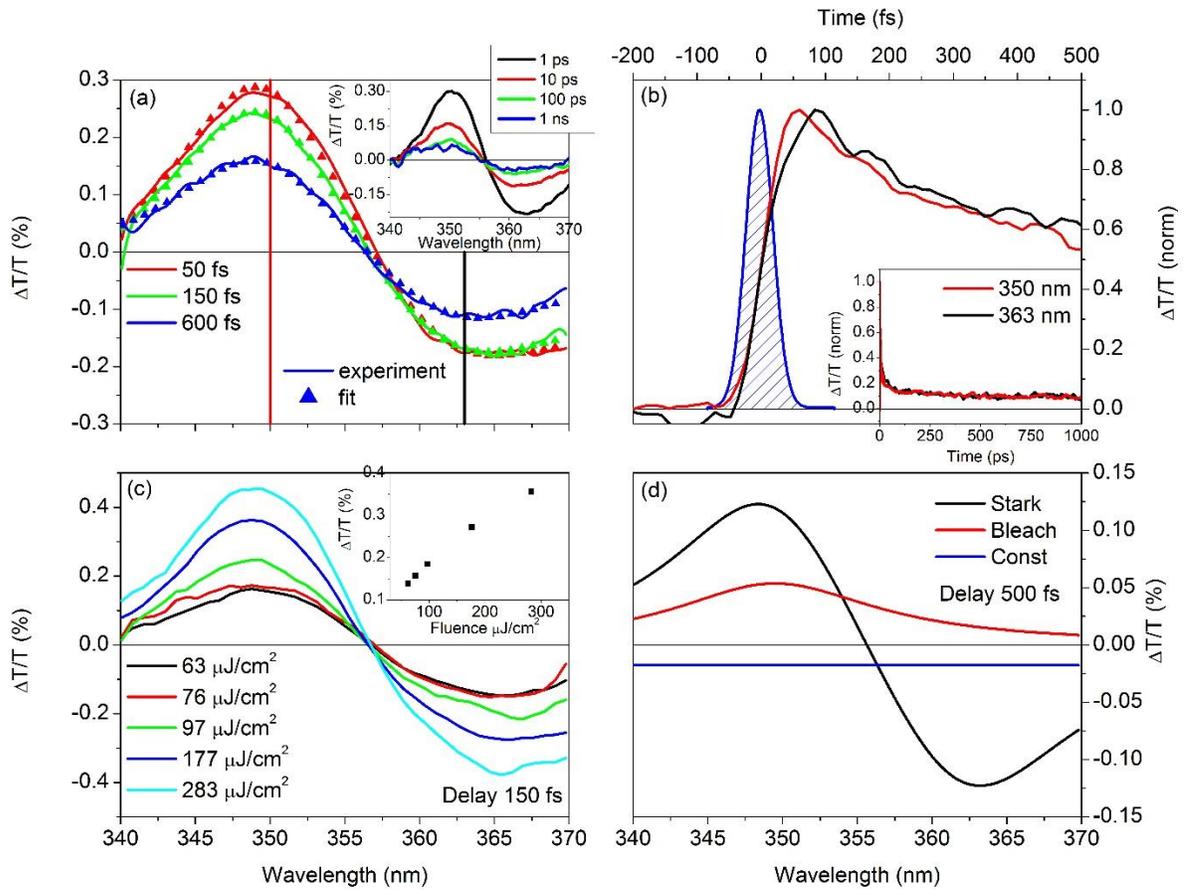

Figure 2.

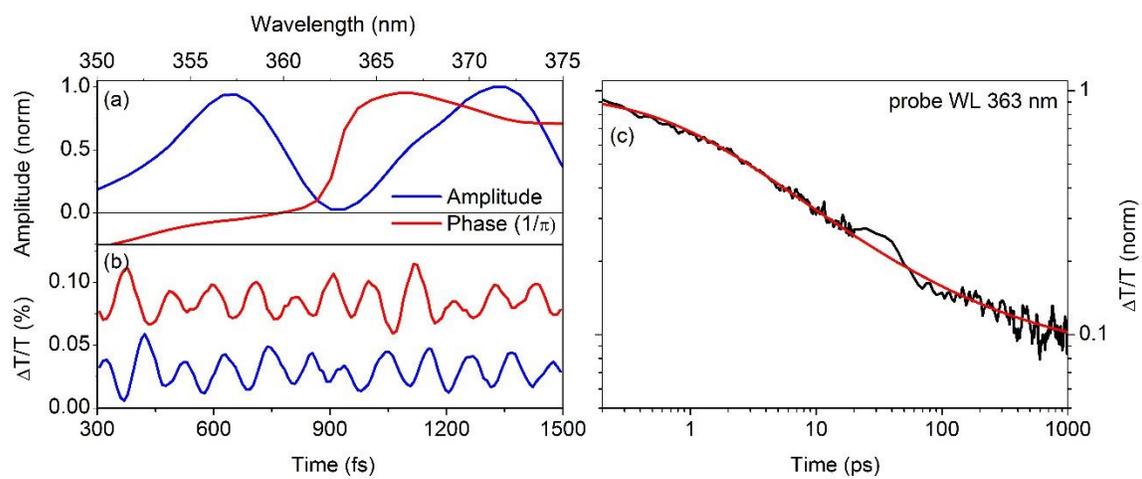

Figure 3.

# High energetic excitons in carbon nanotubes directly probe charge-carriers

Giancarlo Soavi[1], Francesco Scotognella,[1,2,3], Daniele Viola[1], Timo Hefner[4], Tobias Hertel[4], Giulio Cerullo [1, 2*] and Guglielmo Lanzani[1, 3†]

[1] *Dipartimento di Fisica, Politecnico di Milano, Piazza L. da Vinci 32, 20133 Milano, Italy*

[2] *IFN-CNR, Piazza L. da Vinci, 32, 20133 Milano, Italy*

[3] *Center for Nano Science and Technology@PoliMi, Istituto Italiano di Tecnologia, Via Giovanni Pascoli, 70/3, 20133 Milano, Italy*

[4] *Inst. for Physical and Theoretical Chemistry Dept. of Chemistry and Pharmacy, University of Wuerzburg, Wuerzburg 97074, Germany*

## S1. Interpretation of the $\Delta T/T$ signal, temperature and pump photon energy dependence

Although it is commonly agreed that the first derivative of the ground state absorption spectrum can reproduce many of the features observed in the $\Delta T/T$ spectra of semiconducting SWNTs, the physical origin of this peculiar signal is still matter of debate. From a general point of view, a first derivative of the absorption spectrum can be ascribed to a shift of the excitonic transitions. This consideration can only exclude excited state absorption (ESA) as a possible origin of the observed signal, considering that many of the observed photoinduced absorption (PA) bands do not match with any possible transition from excited states [1]. The most evident demonstration of this assumption is indeed the $\Delta T/T$ signal of $S_{33}$, where a sharp PA band appears at approximately 360 nm (≈ 3.44 eV). If we consider that, after initial relaxation processes, all the photoexcited population is on $S_{11}$, thus at approximately 1 μm (≈ 1.24 eV), the observed ESA transition would end up in a state at approximately 265 nm (≈ 4.68 eV). Although we can definitely exclude ESA, the derivative shape in the $\Delta T/T$ signal can arise from many different photoexcited species, ranging from bi-excitons [2] to trions [3], thermal effects [4] and Stark effect [5].

To exclude thermal effects we performed temperature dependent measurements using a cryostat equipped with a liquid nitrogen reservoir. We observed that the signal is weakly sensitive to changes in temperature (Fig. S1): a very small red-shift of the zero, approximately 2 nm when moving from 300 K to 130 K, can be detected for decreasing temperatures. This excludes geometrical re-arrangement (i.e. diameter distortion) and thermal effects as a possible origin of the observed signal, as the increase in temperature is extremely weak in TA experiments at 1 kHz [5] and the $\Delta T/T$ signal is reproduced by a shift of more than 10 nm. To exclude bi-excitons and trions we performed measurements with different pump-photon energies on a very broad probe region, ranging from 340 nm to 650 nm. We notice that: i) we always obtain the same derivative shape both for resonant and non-resonant excitation with respect to the excitonic transitions (Fig. S2); ii) we observe a clear derivative shape in a probe region far from any excitonic peak (Fig. S2). The first observation (i) rules out bi-excitons as a possible explanation, as we expect that more excitons, and consequently bi-excitons, are formed with resonant excitation. Trions, instead, can arise when a charge is photogenerated from the pump pulse and thus can also appear with non-resonant excitation. The second observation (ii) points out that the whole ground state absorption spectrum (and not only excitons) undergoes a red shift upon photoexcitation. This excludes both bi-excitons and trions, whose appearance should depend selectively on the excitonic transitions. On the contrary, the Stark effect lacks this selectivity and simply modifies the whole absorption spectrum.

Indeed, the interpretation of the derivative shape in the $\Delta T/T$ as induced by Stark effect can be successfully extended to other modulation spectroscopy experiments. For example, from the analysis of the signal on $S_{33}$ it appears that the presence of a dense continuum of states plays a fundamental role in the $\Delta T/T$ spectra and thus we expect to find a similar derivative shape already on $S_{11}$ for those tubes with small binding energy, such as metallic tubes. This experiment was performed on isolated metallic CNTs by Gao *et al*. [6], resulting in the expected derivative shape which was, instead, interpreted in terms of bi-excitons. Similarly, Ham *et al.* [7] found a surprisingly high electro-absorption signal in the UV region with respect to the first and second excitonic subbands, which was further enhanced by the presence of metallic CNTs. This discussion deserves at least two additional comments. First, the red-shift of the ground state absorption spectrum induced by bi-excitons or trions is determined by their binding energy and thus we would expect a bigger red-shift for strongly bound excitons [8] such as $S_{11}$ with respect to $S_{33}$ or, similarly, for semiconducting SWNTs with respect to metallic ones. On the contrary, the red-shift due to Stark effect is inversely proportional to the exciton binding energy [9, 10], in accordance with the experimental evidences. These considerations further exclude bi-excitons and trions as a possible explanation for the observed $\Delta T/T$ signal. Second, the nature of $S_{33}$ in semiconducting SWNTs is still matter of debate. Rayleigh scattering experiments have demonstrated that it is consistent with an excitonic model [11], nevertheless it shows features ascribable to unbound electron-hole pairs [12, 13], in contrast with theoretical studies which predict high binding energies [14]. Our results confirm that $S_{33}$ has a low binding energy, in analogy with $M_{11}$ of metallic tubes.

Finally, we check that the power law decay dynamics of the 363 nm signal is independent from the excitation fluence, thus excluding bimolecular recombination processes (Fig. S3).

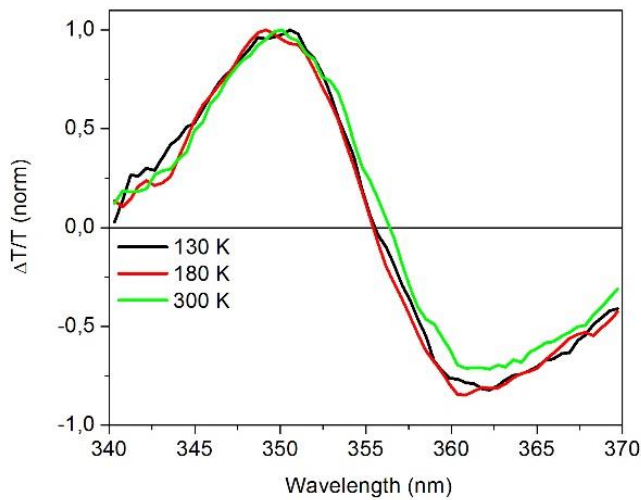

**Figure S1.** $\Delta T/T$ spectra at a pump-probe delay of 10 ps as a function of temperature. The sample was excited with a broadband IR pulse while the probe was obtained by white light generation on a $CaF_2$ plate.

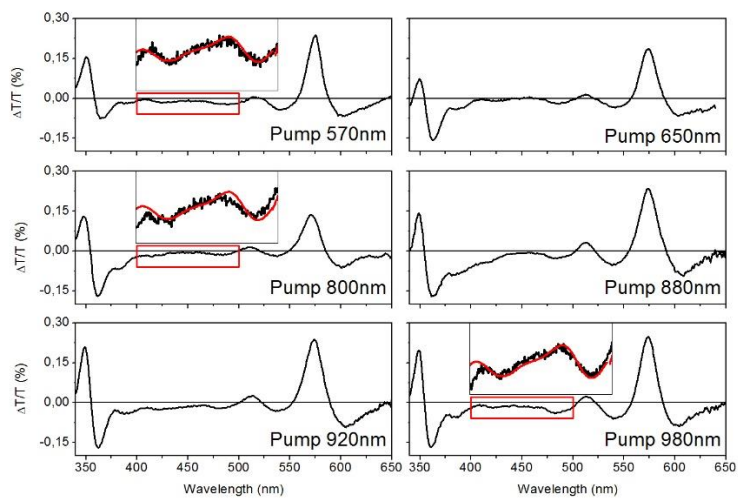

**Figure S2.** $\Delta T/T$ spectra at a pump-probe delay of 30 ps for six different pump-photon energies, resonant and non-resonant with respect to the excitonic transitions of the (6,5) CNT sample under investigation. Here each pump pulse had 10 nm bandwidth while the probe was obtained by white light generation on a $CaF_2$ plate. For pump wavelengths of 570 nm, 800 nm and 980 nm we zoom the probe region between $S_{22}$ and $S_{33}$ (from 400 nm to 500 nm) and we fit it with the $\Delta T/T$ obtained from a 100 meV shift of the ground state absorption spectrum[5].

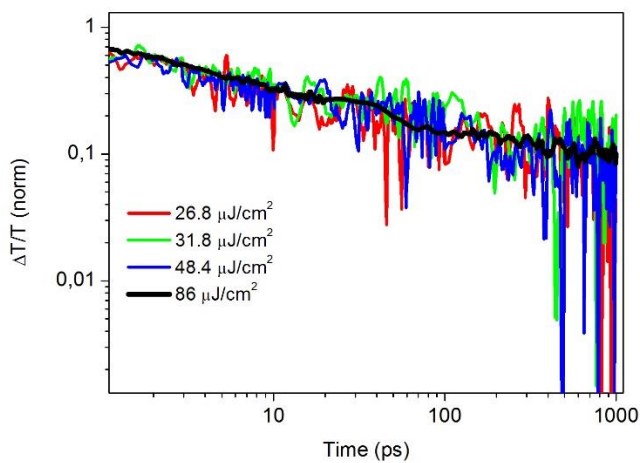

**Figure S3.** Dynamics at 363 nm probe wavelength and broadband IR excitation for different pump fluences.

## S2. Fitting procedure

In Fig. S4 we compare experimental results with our fitting model. For any pump-probe delay we calculate the ΔA transient spectra from $\Delta T/T$, with $A = -\ln(T)$, as a function of the probe energy (instead of wavelength). This procedure is useful in order to have a direct comparison with the O.D. of the ground state absorption spectrum. Each spectrum (namely each ΔA at a fixed pump-probe delay) is reproduced by the sum of three contributions: $\Delta A_{fit} = \Delta A_{bleach} + \Delta A_{Stark} + K$. First, we run the fitting routine for each ΔA spectrum (i.e. at different pump-probe delays) and we set the shape (i.e. peak position and broadening) of $\Delta A_{bleach}$ and $\Delta A_{Stark}$ from the average of the values obtained for each pump-probe delay. After this, the shape is kept constant and we only vary its amplitude as a function of the pump-probe delay. We thus obtain both spectral information and dynamics (the evolution of the amplitude of each component) of the three signals. We decided to keep the shape of $\Delta A_{bleach}$ and $\Delta A_{Stark}$ fixed since, from the first fitting routine, the parameters undergo only slight changes without any specific temporal trend. In particular: i) $\Delta A_{bleach}$ is obtained from a Lorentzian function, typical of excitonic transitions [11], peaked at 3.55 eV with 170 meV FWHM; ii) $\Delta A_{stark}$ is the difference between two Lorentzian functions with 170 meV FWHM peaked at 3.42 eV and 3.55 eV; iii) $K$ is a constant over the entire probe energy region. Interestingly the constant $K$, which we attributed to excited state absorption to the continuum of states [5], shows exactly the same dynamics of the PB signal, indicating that it arises from $S_{11}$. This PA band was already observed at low energies (≈ 300-400 meV), setting a lower limit for the exciton binding energy [15].

Finally, we convert again our model to $\Delta T/T$ as a function of the probe wavelength. The large broadening of the Lorentzian excitonic function for high energetic excitons is in good agreement with recent experiments [16] and might be additionally altered by environmental effects [17]. This kind of analysis is necessary in order to separate the different contributions arising from excitons (PB and PA) and charge-carriers (Stark effect). From the study of the obtained dynamics (Fig. S4), we observe that the three components are formed within our temporal resolution (≈ 50-fs). The linear trend of the signal with respect to the pump fluence excludes non-linear processes such as two-photon absorption or exciton-exciton annihilation [3]. Thus, the most feasible mechanisms for charge-carrier photogeneration remain direct excitation or ultrafast linear exciton dissociation. Direct excitation is easily accessible with high energetic photons [9] while possible mechanisms for exciton dissociation are still largely discussed. Recent experiments [18] show that $S_{22}$ will more likely undergo dissociation into free electron-hole pairs with respect to relaxation into $S_{11}$. Nevertheless, $S_{22}$ lies in the continuum of states and thus exciton dissociation processes are more likely, while dissociation of $S_{11}$ was predicted to occur only in presence of intense external electric fields [19]. Experiments in this direction are indeed controversial, showing both field-induced exciton dissociation [20, 21] or linear exciton dissociation and instantaneous free carrier generation [5, 22, 23].

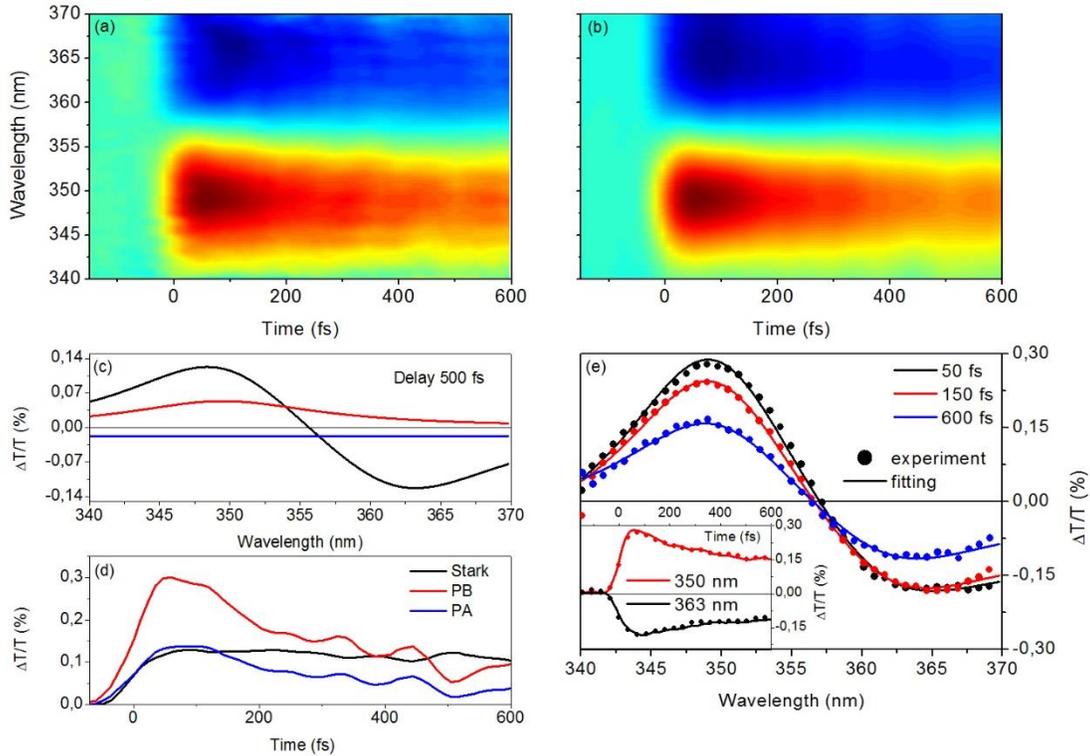

**Figure S4.** Comparison between experimental (a) and fitted (b) $\Delta T/T$ map. (c) Spectra and their amplitude/ dynamics (d) (inset) used for the fit. (e) Comparison between the fit and experimental data.

### S3. Bi-excitons and Stark effect

Both bi-exciton and Stark shift are Coulomb related phenomena, and bear similarity. For this reason it is worth to further discuss the experimental data in order to confirm that the observed red-shift of the absorption spectrum actually arises from Stark effect.

1. **Different dynamics between $S_{11}$ or $S_{22}$ and $S_{33}$**. The interpretation of the pump-probe signal as a bi-exciton implies that that the decay dynamics are the same for the $S_{11}$, $S_{22}$ and $S_{33}$ transitions, given that the excitation conditions are the same. In particular, they will follow the decay dynamics of excitons since the pump pulse will only generate excitons and their lifetime will determine the probability for the probe pulse to generate a bi-exciton. This is not what we observe in our experiments (Fig. S5). When we compare the decay dynamics at the peak of the bleaching for $S_{22}$ (570 nm) and $S_{33}$ (350 nm) under *exactly the same experimental conditions* (broadband IR pump at resonance with $S_{11}$ and simultaneous probe of $S_{22}$ and $S_{33}$ with white light generated in $CaF_2$), we find a faster decay for $S_{22}$ compared to $S_{33}$ (Fig. S5.a). A power law decay dynamics was observed also for the $S_{11}$ excitonic transition in ensembles of semiconducting SWNT [24, 25] after approximately 10 ps. This fully supports our interpretation since $S_{11}$, according to our model, will decay as others ($t^{-0.5}$), once excitonic features are decayed. Moreover, in Ref. 24 and 25 due to spectral congestion, detailed assignment of the dynamics is very difficult and spectral overlap may affect the kinetics. The observed dynamics were assigned to either bimolecular triplet [24] or singlet [25] annihilation. These assignments are in contrast with our work, which shows for the first time a diffusion limited geminate recombination in a single chirality specimen and in the linear excitation regime. Our analysis of the spectral shape of the $S_{33}$ ΔT/T signal allows to assign the transient signal to charge carrier recombination, enlightening previous results. Charge carriers are never mentioned in Ref. 24 and 25 but our work clarifies that charge carriers were indeed involved in those cases as well.

2. **Derivative shape is independent from excitation wavelength**. we have shown that the derivative shape over a very broad probe wavelength region is independent from the pump wavelength (also in agreement with Ref. 22). The observed energy shift depends on the local field that in turns depends on the local geometry, but enhancing the number of charges leads to more modulation sites, not to a larger modulation. In the experiments, the fluence for each excitation wavelength was adjusted in order to obtain the same transient absorption signal. Nevertheless, since we study the shape of the ΔT/T signal (the energy-shift) and not its amplitude, the number of photogenerated charges is not of primary importance. To explain this in term of bi-excitons, instead, we need to assume that any excitation wavelength leads to the formation of $S_{11}$ excitons but instead we know that: i) $S_{11}$ excitation generates <u>also</u> charges [5, 22]; ii) $S_{22}$ excitation generates charges with high yield due to spontaneous exciton dissociation [18]; iii) above-gap out of resonance excitation does not give rise to excitons, but on the other hand it has been shown to generate charges.

3. **Red-shift affects also a spectral region far from excitons**. In our previous work [5] we showed that the red-shift of the absorption spectrum affects also a spectral region far from excitonic transitions. This can be observed also in Figure S2, where we zoom in on the region between 400 nm and 500 nm and we show (red line) the fit obtained by a simple red-shift of the absorption spectrum. In this probe region there are no excitonic transitions related to the (6,5) SWNTs and thus the signal cannot be attributed to bi-excitons. These spectral features can be easily interpreted in terms of charge-induced Stark effect since the induced electric field lacks selectivity and acts on the whole absorption spectrum.

4. **Binding energy of the biexciton is not consistent**. In the case of bi-exciton the red-shift of the absorption spectrum can be used to derive its binding energy [26] and we expect that a bi-exciton formed by $S_{11}+S_{22}$ ("excited bi-exciton") is more stable than a $S_{11}+S_{33}$ bi-exciton. This would result in larger red-shifts on $S_{22}$ with respect to $S_{33}$, in contrast with our experimental evidences that show exactly the opposite trend. In fact the shift on $S_{22}$ is always smaller than the value obtained for $S_{33}$. Considering the calculation by Pedersen *et al.* [27] for the $S_{11}$-$S_{11}$ bi-exciton we can estimate that $E_{12} > E_{13}$ (where $E_{xy} = E_x +$ $E_y - E$ is the binding energy of the excited bi-exciton, $E_x$ and $E_y$ are the binding energies of the single excitons and E is a positive constant) if we assume that $E_2 > E_3$ [12]. On the other hand, the Stark effect is consistent with a larger red-shift on $S_{33}$ with respect to $S_{22}$ since it scales like the inverse of the exciton binding energy.

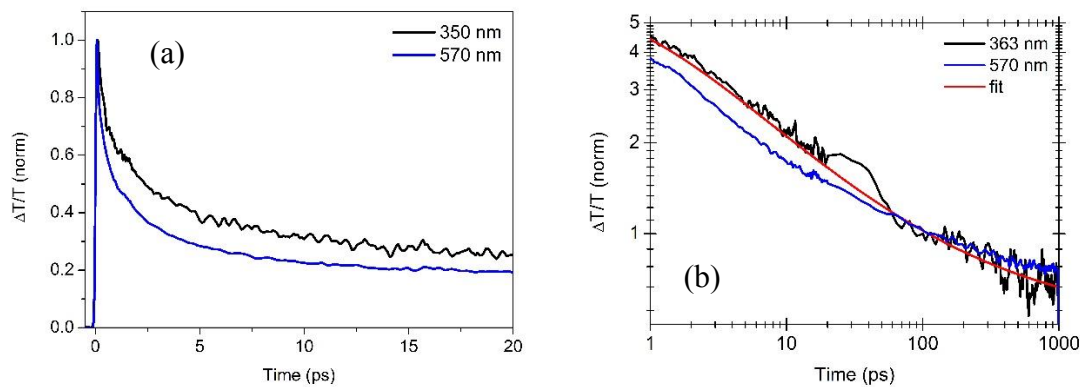

**Figure S5.** a) Normalized decay dynamics of the $S_{33}$ (350 nm, black line) and $S_{22}$ (570 nm, blue line) excitons for broadband IR excitation, at resonance with $S_{11}$. b) Decay dynamics of the $S_{33}$ (350 nm, black line) and $S_{22}$ (570 nm, blue line) excitons normalized at 100 ps for broadband IR excitation, at resonance with $S_{11}$.

**Supplementary References**

[1] B. Gao, G. V. Hartland and L. Huang, ACS nano **6**, 5083 (2012).

[2] D. J. Styers-Barnett, S. P. Ellison, B. P. Mehl, B. C. Westlake, R. L. House, C. Park, K. E. Wise and J.   M. Papanikolas J. Phys. Chem. C **112**, 4507 (2008).

[3] S. M. Santos, B. Yuma, S. Berciaud, J. Shaver, M. Gallart, P. Gilliot, L. Cognet and B. Lounis Phys. Rev. Lett. **107**, 187401 (2011).

[4] T. Koyama, S. Yoshimitsu, Y. Miyata, H. Shinohara, H. Kishida and A. Nakamura, J. Phys. Chem. C   **117**, 20289 (2013).

[5] G. Soavi, F. Scotognella, D. Brida, T. Hefner, F. Späth, M. R. Antognazza, T. Hertel, G. Lanzani and   G. Cerullo, J. Phys. Chem. C **117**, 10849 (2013).

[6] B. Gao, G. V. Hartland and L. Huang, J. Phys. Chem. Lett. **4**, 3050 (2013).

[7] M.-H. Ham, B.-S. Kong, W.-J. Kim, H.-T. Jung and M. S. Strano, Phys. Rev. Lett. **102**, 047402   (2009).

[8] D. Kammerlander, D. Prezzi, G. Goldoni, E. Molinari and U. Hohenester, Phys. Rev. Lett. **99**, 126806   (2007).

[9] J. Crochet, S. Hoseinkhani, L. Lüer, T. Hertel, S. K. Doorn and G. Lanzani, Phys. Rev. Lett. **107**,   257402 (2011).

[10] M. S. Arnold, J. L. Blackburn, J. Crochet, S. K. Doorn, J. G. Duque, A. Mohitee and H. Telg, Phys. Chem. Chem. Phys. **15**, 14896 (2013).

[11] S. Berciaud, C. Voisin, H. Yan, B. Chandra, R. Caldwell, Y. Shan, L. E. Brus, J. Hone and T. F. Heinz, Phys. Rev. B **81**, 041414 (2010).

[12] P. T. Araujo, S. K. Doorn, S. Kilina, S. Tretiak, E. Einarsson, S. Maruyama, H. Chacham, M. A. Pimenta and A. Jorio, Phys. Rev. Lett. **98**, 067401 (2007).

[13] T. Michel, M. Paillet, J. C. Meyer, N. V. Popov, L. Henrard and J.-L. Sauvajol, Phys. Rev. B **75**, 155432 (2007).

[14] C. D. Spataru, S. Ismail-Beigi, L. X. Benedict and S. G. Louie, Phys. Rev. Lett. **92**, 077402 (2004).

[15] Y.-Z. Ma, L. Valkunas, S. M. Bachilo and G. R. Fleming, J. Phys. Chem. B **109**, 15671 (2005).

[16] E. H. Haroz, S. M. Bachilo, R. B. Weisman and S. K. Doorn, Phys. Rev. B **77**, 125405 (2008).

[17] J.-C. Blancon, M. Paillet, H. N. Tran, X. T. Than, S. Aberra Guebrou, A. Ayari, A. San-Miguel, N.- M. Phan, A.-A. Zahab, J.-L. Sauvajol *et al.*, Nat. Commun. **4**, 2542 (2013).

[18] Y. Kumamoto, M. Yoshida, A. Ishii, A. Yokoyama, T. Shimada and Y. K. Kato, Phys. Rev. Lett. **112**, 117401 (2013).

[19] V. Perebeinos and P. Avouris, Nano Lett. **7**, 609 (2007).

[20] A. D. Mohite, P. Gopinath, H. M. Shah and B. W. Alphenaar, Nano Lett. **8**, 142 (2008).

[21] A. Mohite, J.-T. Lin, G. Sumanasekera and B. W. Alphenaar, Nano Lett. **6**, 1369 (2006).

[22] M. C. Beard, J. L. Blackburn and M. J. Heben, Nano Lett. **8**, 4238 (2008).

[23] S. A. Jensen, R. Ulbricht, A. Narita, X. Feng, K. Müllen, T. Hertel, D. Turchinovich and M.Bonn, Nano Lett. **13**, 5925 (2013).

[24] R. Russo, E. Mele, C. Kane, I. Rubtsov, M. Therien and D. Luzzi, Phys. Rev. B **74**, 041405(R) (2006).

[25] J. Allam, M. T. Sajjad, R. Sutton, K. Litvinenko, Z. Wang, S. Siddique, H. Q. Yang, W. H. Loh and T. Brown, Phys. Rev. Lett **111**, 197401 (2013).

[26] V. I. Klimov, Annu. Rev. Phys. Chem. **58**, 635 (2007).

[27] T. G. Pedersen, K. Pedersen, H. D. Cornean and P. Duclos, Nano Lett. **5**, 291 (2005).